# Photogrammetry and ballistic analysis of a high-flying projectile in the STS-124 space shuttle launch

Philip T. Metzger
NASA/KSC Granular Mechanics and
Regolith Operations Lab, KT-D3
Kennedy Space Center, FL 32899
Philip.T.Metzger@nasa.gov

John E. Lane
ASRC Aerospace, ASRC-24
Kennedy Space Center, FL 32899
John.E.Lane@nasa.gov

Robert A. Carilli
United Space Alliance, USK-751
Kennedy Space Center, FL 32899
Robert.A.Carilli@nasa.gov

Jason M. Long
ASRC Aerospace
4231 Suitland Road
Suitland, MD 20726
Jason.Long@noaa.gov

Kathy L. Shawn
United Space Alliance, USK-503
Kennedy Space Center, FL 32899
Kathy.L.Shawn@nasa.gov

**Abstract**

A method combining photogrammetry with ballistic analysis is demonstrated to identify flying debris in a rocket launch environment. Debris traveling near the STS-124 Space Shuttle was captured on cameras viewing the launch pad within the first few seconds after launch. One particular piece of debris caught the attention of investigators studying the release of flame trench fire bricks because its high trajectory could indicate a flight risk to the Space Shuttle. Digitized images from two pad perimeter high-speed 16-mm film cameras were processed using photogrammetry software based on a multi-parameter optimization technique. Reference points in the image were found from 3D CAD models of the launch pad and from surveyed points on the pad. The three-dimensional reference points were matched to the equivalent two-dimensional camera projections by optimizing the camera model parameters using a gradient search optimization technique. Using this method of solving the triangulation problem, the xyz position of the object's path relative to the reference point coordinate system was found for every set of synchronized images. This trajectory was then compared to a predicted trajectory while performing regression analysis on the ballistic coefficient and other parameters. This identified, with a high degree of confidence, the object's material density and thus its probable origin within the launch pad environment. Future extensions of this methodology may make it possible to diagnose the underlying causes of debris-releasing events in near-real time, thus improving flight safety.

*Keywords:* Photogrammetry; Coefficient of drag; Launch debris; Ballistic coefficient; Pin-hole camera model; Launch Pad 39A

# 1. Introduction

An anomalous debris event occurred during the launch of Space Shuttle Discovery on STS-124, May 31, 2008. Several thousand bricks were unexpectedly released from the east wall of the launch pad flame trench, on the north end, which channels the Solid Rocket Booster (SRB) exhaust plumes. These bricks were observed to blow in a largely horizontal and northward direction, and they were found deposited over a wide area north of the launch pad. Most of the bricks were broken into small fragments and landed in a zone that is adjacent to the launch pad and reaches to its perimeter fence. Many of these lay just inside the fence, presumably due to their impact and rebound from the fence, which was severely damaged by the large number of impacts. Other bricks or fragments of bricks were observed splashing into the marsh waters north of the launch pad, just outside the perimeter fence, presumably flying not-too-high over the top of the fence. If all the bricks and fragments blew in this nearly horizontal direction toward the north, then there was no safety concern for the mission.

However, simultaneous with this event, several unidentified projectiles were observed flying in a different direction: vertically away from the base of the Space Shuttle, Mobile Launch Platform (MLP), and/or flame trench, while still on the north side of the vehicle. This was a significant concern for future Space Shuttle launches because, if these were brick fragments blown into vertical trajectories, then we must conclude that debris from beneath the MLP had found a pathway to enter the space above the MLP, where they could in theory strike the launch vehicle. It has always been assumed that debris does not blow from beneath the MLP to the space above it. On the other hand, these vertical projectiles might not have been brick fragments. There are several known sources of material that are blown routinely during Space Shuttle launches. These include the SRB throat plug foam (a low-density foam that blows out from the SRB throat at ignition and before ignition serves to seal the SRBs against insects, birds, or other intrusions) and the SRB water bags (cloth bags filled with water to be released at ignition to aid in dampening the exhaust plume transient effects). Other than these expected materials and the anomalously released bricks, sources of projectiles might include debris that was erroneously left unsecured at the launch pad and/or materials that were broken loose by the high-energy effects of launch. It was therefore important to determine whether these vertical projectiles, or "high-flyers" as they came to be called, were associated with the brick release event or another anomalous condition or whether they were a part of the ordinary release of known materials. Even if these projectiles came from known sources, it is very important to identify and understand their trajectories to quantify the risk they might impose to the launch vehicle.

This report makes no attempt to provide an integrated analysis of these risks. Rather, it describes one contribution to an overall analysis. It focuses on the trajectory and the material identity of one of the high-flyers, demonstrating the analysis technique that was developed for this problem. We show that this high-flyer originated from the same location as the anomalously released flame trench bricks (the east wall of the north end of the flame trench, at the same approximate location as the flame bricks). Surprisingly, however, the high-flyer was not made of brick material. Rather, we show that it was made of a material having the same density as SRB throat plug foam. Two possible explanations for this unexpected result are proposed in Section 5.

## 2. Photogrammetry analysis

*2.1 Pinhole camera model*

The ideal camera model is based on the pinhole camera model (PCM), as shown in Fig. 1. The focal length of a pinhole camera is simply the distance from the pinhole to the image plane. Practical characteristics of real cameras can be modeled by the PCM. Since an ideal pinhole has a diameter of zero, the object image projected on the image plane has zero intensity. The job of a lens system is to allow a large camera aperture to behave as a pinhole, thus providing useful light intensity at the image plane. Differences between the optical characteristics of a pinhole camera and a camera with a lens can be defined to be due to lens distortion, since by design, an ideal lens can be defined to behave as a pinhole with a finite aperture. In photogrammetry applications, an unknown distance in space $\Delta x$ can be determined by measuring the equivalent image distance $\Delta u$, so that $\Delta x = \Delta u \, (z - f)/f$, as shown in Fig. 1.

Neglecting lens distortion, all characteristics of the spatial transformation of three-dimensional world coordinate $(x, y, z)$ to a two-dimensional image coordinate $(m, n)$ can be modeled by the PCM of Fig. 1. There are numerous ways to extend the PCM to a general three-dimensional model of the camera, relating various physical parameters to the camera model. For example, the camera model described by Equation (6.7) of Hartely & Zisserman (H&Z) [1], p.156 of the 2$^{nd}$ edition, is a generalized 13-parameter model.

The debris trajectory analysis software developed at the Kennedy Space Center (KSC) was originally developed during a 2003 accident investigation and, because of the nature of the 'scene' and camera images, a far-field (affine) model was used. The image analysis algorithms utilized by participants of that accident investigation were mostly CAD-based approaches and not strictly photogrammetry methods [2, 3]. The KSC trajectory analysis software [4] was expanded to accommodate near field scenes such as those that are the focus of this paper, by modifying the affine camera model to include the effects of a pin-hole camera, as shown in Equation (1). Note that by setting $R_x = R_y = R_z = 0$ in Equation (1), the camera model is reduced to the original affine model that was developed to support the Columbia Accident Investigation Board (CAIB) report. The photogrammetry analysis performed in this work is based on the following 11-parameter camera model:

$$\begin{pmatrix} m \\ n \end{pmatrix} = \left\{ (R_x \ R_y \ R_z) \cdot \begin{pmatrix} x \\ y \\ z \end{pmatrix} + 1 \right\}^{-1} \left\{ \begin{pmatrix} a_x & b_x & c_x \\ a_y & b_y & c_y \end{pmatrix} \cdot \begin{pmatrix} x \\ y \\ z \end{pmatrix} + \begin{pmatrix} m_0 \\ n_0 \end{pmatrix} \right\} + \begin{pmatrix} m_c \\ n_c \end{pmatrix}, \qquad (1)$$

where $m$ and $n$ are the image pixel coordinates of a projected point at $x$, $y$, and $z$ in world coordinates, and the unknown camera parameters are $a_x$, $b_x$, $c_x$, $a_y$, $b_y$, $c_y$, $R_x$, $R_y$, $R_z$, $m_0$, and $n_0$. The constants $(m_c, n_c)$ are the image coordinates of the camera optical axis and are set to the center pixel coordinates of the image. This simplification only applies to cases where it can be legitimately assumed that the center of the charged-coupled device (CCD) and camera optical axis are properly aligned.

The camera model described by Equation (1) can be shown to be equivalent to the camera model described by Equation (6.7) of Hartely & Zisserman (H&Z). The parameters $a_x$, $b_x$, $c_x$, $a_y$, $b_y$, and $c_y$, in Equation (1) are equivalent to the first two rows of the Euler rotation matrix, multiplied by the focal length $f$, divided by the dot product of a vector formed by the third row of the rotation matrix and the negative of the vector **C** in Equation (6.7) of H&Z. The parameters $R_x$, $R_y$, and $R_z$ are equivalent to the third row of the rotation matrix, divided by the same dot product described above. The parameters $m_0$ and $n_0$ are equivalent to the product of the negative of the 3x2 matrix in Equation (1) and vector **C** from Equation (6.7) of H&Z.

*2.2 Determination of camera model parameters*

An error function $E$ can be formed from the camera model, Equation (1), using $N$ reference points visible in the camera image. Each of the $k$th reference points consists of five known values: three world coordinates, $(x_k, y_k, z_k)$ and two image pixel coordinates, $(m_k, n_k)$. The error function $E$ is a function of the 11 camera parameters and is formed by the following sum:

$$E = E(a_x, b_x, c_x, a_y, b_y, c_y, R_x, R_y, R_z, m_0, n_0)$$

$$= \sum_{k=1}^{N} \left\{ \begin{pmatrix} \mu_k \\ \nu_k \end{pmatrix} (R_x x_k + R_y y_k + R_z z_k + 1) - \begin{pmatrix} a_x & b_x & c_x \\ a_y & b_y & c_y \end{pmatrix} \cdot \begin{pmatrix} x_k \\ y_k \\ z_k \end{pmatrix} + \begin{pmatrix} m_0 \\ n_0 \end{pmatrix} \right\}^2, \quad (2)$$

where $\mu_k \equiv m_k - m_c$ and $\nu_k \equiv n_k - n_c$. Finding the minimum of $E$ provides an estimate of the optimum set of 11 parameters required by Equation (1). Appendix A describes the method of minimizing $E$ used by the KSC photogrammetry software. The matrix $\Gamma$ in Equations (A-4a and A-4b) is computed as the sum of all outer products of $\alpha_k$ with itself plus the outer product of $\beta_k$ with itself.

Fig. 2 depicts the location of the highspeed film cameras, E060 and E062, used in this analysis. Fig. 3 shows the view of reference points corresponding to cameras E060 and E062. These reference points were provided by a differential GPS survey, accurate to better than one mm. Viewable locations on the fixed service structure (FSS) and MLP, as well as on the water tower, were surveyed as photogrammetry reference points. Table 1 lists the coordinates of the surveyed reference points, corresponding to Fig 3. Note that the SE corner of the MLP is chosen as the coordinate system origin, where north is $x$, west is $y$, and up is $z$.

*2.3 Combining multiple camera data*

Now consider multiple cameras viewing the same object at the unknown position $(x, y, z)$. Synchronized (or temporally interpolated) image data from two or more cameras can be combined, as in the previous section, using an error minimization strategy. In this case, an error function $E = E(x, y, z)$ can be defined by summing the difference terms, analogous to Equation (2), for all $M$ cameras (for the specific case of this analysis, $M = 2$):

$$E = E(x, y, z)$$

$$= \sum_{j=1}^{M} \left\{ \begin{pmatrix} \mu_j \\ v_j \end{pmatrix} (R_{x_j} x + R_{y_j} y + R_{z_j} z + 1) - \begin{pmatrix} a_{x_j} & b_{x_j} & c_{x_j} \\ a_{y_j} & b_{y_j} & c_{y_j} \end{pmatrix} \begin{pmatrix} x \\ y \\ z \end{pmatrix} + \begin{pmatrix} m_{0_j} \\ n_{0_j} \end{pmatrix} \right\}^2. \qquad (3)$$

The unknown coordinates can be found by setting the gradient of E to zero and finding the solution, as described in Appendix B. Fig. 4 shows the reference points and debris track found using the procedures described above. Fig. 5 is a close-up of the debris observed in frames 6-8. Note that one pixel is roughly equal to 1 inch. The top and bottom markers are used to estimate the debris size, as shown by the graph of Fig. 6. Trajectory data points are given in Table 2. These are reported in inches to agree with the coordinate systems used in the Space Shuttle program.

### 3. Ballistic Analysis Methods

To calculate drag forces on the projectile, it will be necessary to know its size. For the present problem in distinguishing brick from foam, only rough estimates should be needed. However, we sought to develop the best possible estimates taking into consideration the nonsphericity of the object in order to evaluate the effectiveness of the generic methodology. The projectile appears with fuzzy edges in the photographs and so its size can only be estimated. It appears to be elongated in one axis, perhaps with a roughly ellipsoidal shape. The outer extremes of the fuzzy edges have dimensions of 19 cm in the long dimension and 11.5 cm in the shorter dimension. The third dimension is presumably about the same size as the second dimension as no shorter or longer dimension appears in the tumbling. Blurred objects are usually smaller than they appear in an image, so it is reasonable to scale these dimensions down slightly to their original size. We estimated these to be 15.25 cm (long dimension) and 9 cm (both shorter dimensions). We did not have detailed drag force data for rotating ellipsoids, so we used an "equivalent sphere" approximation. We found the diameter of a sphere that has the same cross-sectional area (projected into the direction of travel) as the time-averaged cross-sectional area of this tumbling ellipsoid. For the projectile with major axis = 15.25 cm and both minor axes = 9 cm, it can be shown that $D_{equiv} = 10.86$ cm.

Because the projectile is in ambient air, it should have negligible crossrange acceleration, and thus the trajectory should exist in a single, vertical plane that is at some angle relative to the $x$ and $y$ axes. The first analysis step is to reduce the number of parameters by transforming the coordinates into this plane. Linear regression of the $(x, y)$ points is shown in Fig. 7.

The data points do not lie perfectly in the plane. This may be due to photogrammetric errors or due to the aerodynamics of the projectile providing horizontal force. The motion out of the plane as shown in Fig. 7 appears to be smooth, not random, but it is sufficiently small that it will be neglected. The distance in-plane from the origin is calculated trigonometrically. The reduced data set, converted to metric units, is shown in Table 3. Components of velocity due to out-of-plane motion are neglected, so the velocities are the same as in Table 1.

To perform the ballistic analysis on these data, we wrote time-integration algorithms in Mathematica (v.7.0.0, Wolfram Research, Inc., Champaign, Illinois). Numerically integrating the ballistics, rather than obtaining an analytical solution, allows for the coefficient of drag $C_D = C_D(Re, Ma)$ to be variable with Reynolds number $Re$ and Mach number $Ma$, which in general does not admit an analytical solution. It was necessary to implement variable $C_D$ due to the wide range of $Re$. $Ma$ was well below unity in this particular case, but in future cases the effects of compressibility may be significant. We implemented the equation by Loth [5] for coefficient of drag for smooth spheres, which accounts for the full range of $Ma$ and $Re$ to beyond the drag crisis ($Re_{\text{crit}} = 379{,}000$). The drag crisis is when the boundary layer becomes turbulent so that it remains better attached to the sphere, reducing the pressure imbalance across the sphere and thus reducing drag by an order of magnitude. Unfortunately, it turns out that the diameter and velocity measured in our photogrammety produce a $Re$ that would be on the cusp of the drag crisis throughout its trajectory if the projectile had been a smooth sphere. Uncertainty in the exact onset of drag crisis for a nonspherical projectile is important to characterize lest we inject upwards of an order of magnitude error into our drag calculations. While the rougher surface should induce turbulence in the boundary layer at lower $Re$, it is not clear that this will result in a drag crisis at lower $Re$. The very rough surface shape may itself prevent flow attachment despite the turbulence. Indeed, Loth [6] reports that objects with 15% or larger surface irregularities have no drag crisis at all. To bracket the possibilities, we have performed the simulation in two cases: with the value of $Re_{\text{crit}}$ as for a smooth sphere to represent the smooth case, and with $Re_{\text{crit}}$ larger than any $Re$ experienced by the projectile throughout its trajectory (i.e., no drag crisis at all) to represent the very rough, irregular case. We have also simulated cases with a drag crisis at much lower values of $Re_{\text{crit}}$ than for a smooth sphere, but that produced density predictions that are an order of magnitude lighter than SRB foam. Since no such material is known to exist in the launch environment, we can safely rule out that possibility. We have also multiplied Loth's drag equation by a Shape Factor $C_S$ since the projectile is not a sphere. Ordinarily $C_S$ is in the range of 2 to 3 [7] for irregular objects comparing to smooth spheres. We have already taken into account its general elongation in averaging the cross-sectional area of the particle, but not its actual shape or roughness, so a comparable but somewhat smaller $C_S$ may be in order. Hence we have used values of $C_S = 1.5$ and $C_S = 2.0$. We estimate the uncertainty surrounding drag forces for nonspherical objects injects an error on the order of 50% into the final density estimate, although the results below imply that this may be overly pessimistic. The drag equations we have used are plotted in Fig. 8.

The integration used a time-step that is nonadaptive between steps, the value of which is selectable to achieve a desired accuracy. We tested increasingly smaller time-steps until further decreases produced variations in the trajectory that were smaller than some arbitrary norm. Variations in the norm versus time-step were found to follow a power-law, so the norm could be projected to the case of infinitesimal time-steps and thus provide an absolute comparison against any finite time-step. For example, the user could select that the 12-meter-long trajectory be no more than 3.6 mm (0.03%) different in length than if an infinitesimal time-step had been used. A 636 µs was the value selected to achieve 0.03% accuracy in the following analysis for the optimal density.

Since the projectile is outside the plume and is traveling through the ambient atmosphere at relatively low altitude, the drag force and therefore the acceleration is a function of velocity but not of position,

$$\mathbf{a}(t) = \frac{1}{2} \frac{C_D A}{\rho_S V} \rho \, \mathbf{v}(t) \cdot v(t) + \mathbf{g} \; , \tag{4}$$

where $\mathbf{a}$ is the acceleration vector, $C_D$ is coefficient of drag, $A$ is cross-sectional area of the projectile, the mass of the projectile is $\rho_S V$, where $\rho_S$ is the density of the solid material and $V$ is its volume, $\rho$ is the density of air at sea level (for Launch Complex 39 on the day of launch), $\mathbf{v}$ is the velocity vector, and $\mathbf{g}$ is the gravitational acceleration vector. Therefore, the first integration solves future velocity as a function of present acceleration (in terms of present velocity). We implemented the 2nd-order Runge-Kutta [8] method to solve the projectile's velocity as a function of time. The algorithm then performs a second integration to solve for location of the object as a function of time using Verlet integration, still using the acceleration determined in each time-step by the Runge-Kutta method. Since the Verlet algorithm requires knowledge of two prior time-steps to solve each subsequent time-step, the process must be boot-strapped in the first time-step when only the initial state is known. We implemented ordinary Eulerian integration for the first time-step, but using the acceleration determined by the Runge-Kutta method. Since the time-step is small, the error introduced by one time-step in the Eulerian integration is negligible, and all further time-steps use the more accurate integration methods.

This algorithm obtains location as a function of time, which can then be compared to the data points of the projectile's observed trajectory, as quantified by photogrammetry methods. The simulated projectile's density $\rho_S$ and initial position and initial velocity may be adjusted in a simulated annealing process until the predicted trajectory optimally matches the observed trajectory. The end result of the analysis is that we calculate the optimum value of $\rho_S$ and thus determine with high degree of confidence whether the projectile was brick, liberated from the wall of the flame trench (or something of similar high density), or foam, blown out from the throat of the SRB (or something of similar low density), or a material of intermediate density.

To evaluate how well this method predicts the value of $\rho_S$, we used least-squares regression to find the value of $\rho_S$ that best fits the points obtained from the photogrammetry. Since the trajectory calculation is an initial value problem, the regression simultaneously finds the best values for the initial position and initial velocity of the projectile. We allowed the regression to find these initial values for each value of $\rho_S$ independent of the other values of $\rho_S$. That way, each value of $\rho_S$ has an independent chance at proving its ability to fit the data. To simplify the problem, we ignored out-of-plane position of the projectile (as described above) and we assumed that the initial altitude was equal to the altitude of the first data point. The initial position is therefore reduced to one unknown (its downrange component), and the initial velocity is reduced to two unknowns (its downrange and vertical components; no crossrange).

The regression analysis provides $R^2$, the sum of the squares of the residuals, as the norm for "goodness of fit" to the photogrammetry data. The $R^2$ value is calculated for the residuals corresponding to the discrete points at which there is photogrammetry data. The calculated data points for the trajectory were interpolated to obtain points at the precise values of time at which there are photogrammetry points. The square of the distance between calculated and photogrammetric points in real space (downrange and altitude) is the residual for that point. $R^2$ is the sum of these squares. By plotting $\rho_S$ versus $R^2$, we not only see the value of $\rho_S$ that provides the best fit, but we see how narrow is the minima of $R^2$ and thus evaluate the confidence interval on $\rho_S$.

## 4. Ballistic analysis results

Using $C_S = 1.5$, the best fit is for $\rho_S = 52.22$ kg/m$^3$, initial downrange velocity $v_{R0} = -3.20$ m/s, and initial vertical velocity $v_{Z0} = 49.0$ m/s, with a downrange offset in the starting position of $x_0 = -4.6$ m. Since density scales linearly with drag coefficient, we immediately calculate that for $C_S = 2.0$, the best fit is for $\rho_S = 69.63$ kg/m$^3$. Fig. 9 shows the calculated trajectory for this best fit along with the photogrammetry data. The plot compares data points of the numerical simulation and photogrammetry at identical times. It also compares the best result with two cases having (a) much lower and (c) much higher density, each with its remaining parameters optimized to minimize the least squares norm. These two trajectories could have been made to falsely appear much better than they do if we had used only the overall curvature or the final location of each trajectory as the norm. That would have been accomplished by making the initial velocity much higher (for the low-density case) or much lower (for the high-density case) so that the trajectories for every case would end at the same location in space. However, the least squares norm as-formulated here would then have been much worse because the intermediate points along the trajectory would be spaced too far apart or too close together. This illustrates the importance of using the norm as we have formulated it, because it is minimized when the acceleration of the projectile is correct, and it is the acceleration that depends sensitively on projectile density.

To test the sensitivity of the method to other errors, we have reperformed the entire analysis nine times from beginning to end. Each analysis produces a trajectory that is very similar to every other analysis, but offset laterally by as much as 10 meters in some cases due to limitation of repeatability in tagging the reference points (Fig. 3). Fortunately, spatially shifting a trajectory has no effect upon the ballistics analysis. After subtracting out the offsets, it is found that some scatter on the scale of a meter or less remains in the positioning of individual points comparing the trajectories against one another. This scatter is smaller in the vertical dimension compared to the two lateral dimensions, probably because every camera has a direct view of the vertical dimension. Fortunately, errors in the lateral dimension have little effect on the ballistics analysis compared to the vertical dimension. Using $C_S = 1.5$, the set of nine analyses produced a range of best-fitted densities between 54.89 kg/m$^3$ and 55.65 kg/m$^3$ with a mean of 55.18 kg/m$^3$. Thus, compared to the uncertainty arising from the coefficient of drag, there is negligible error anywhere else in the process.

The downrange, crossrange, and vertical errors for each data point are shown in Fig. 10. The errors, especially crossrange and vertical, are seen to be correlated in time. All nine analyses produced essentially identical errors. This implies that the errors are due to real variations in lift and drag of the tumbling of the projectile and/or motion of the

"ambient" air above the flame trench. Since these motions are small compared to the primary movement along the trajectory, they have little effect upon the density calculation.

Fig. 11 shows the calculated $R^2$ norm versus $\rho_S$ for independent regressions performed at each value of $\rho_S$. The narrowness of the minimum indicates the confidence interval in $\rho_S$.

## 5. Identification of Projectile

The flame trench fire bricks have a density of approximately 2260 kg/m$^3$. Their dimensions are approximately 34.29 cm (13.5 in) long, 15.24 cm (6 in) deep, and 7.62 cm (3 in) high. Brick fragments between 1 g in mass up to a half-brick in size were measured near the launch pad and were found to be uniformly distributed at all fragment masses (i.e., the *total* mass of all fragments having a particular mass is the same for every particular mass; thus, there are a vastly higher *number* of fragments at the smaller masses). Thus, it would be possible to find a brick fragment having roughly the dimensions of our projectile, but the density of the brick material is too high by two orders of magnitude. For the projectile to have the density of brick with the $C_D$ and $D_{equiv}$ calculated above, it would correspond to $R^2 = 3.36$, far outside the minimum shown in Fig. 11.

SRB throat plug foam has a density of 48 to 51 kg/m$^3$, consistent with the density calculated by regression of the ballistics. The disk of foam within the SRB throat is 16.5 cm (6.5 in) thick, so fragments will be no larger than this in at least one dimension. Thus we note that it would be possible to find an SRB foam fragment having the dimensions of our projectile. There are very few materials in the launch pad environment with such a low density and fewer still that are likely to be found blowing during a launch. It is highly probable that the projectile was SRB throat plug foam.

If we did not have a small list of potential materials to compare, it would have been more difficult to identify the precise source of the projectiles. However, because the $R^2$ confidence interval is so tight and because the calculated density of the projectile so tightly brackets the known density of foam, we have very high confidence that the projectile has been identified correctly. This is despite the fact that, ironically, the trajectory of the projectile points backward to the same place where bricks were being released from the wall. We do not know why SRB throat plug foam exited the flame trench from the location of the brick release. It may have been deflected by the releasing bricks and the disturbed gas flow around them, or it may be that coming out from under the edge of the Mobile Launch Platform there was sufficient upward convection in the top of the plume to turn the foam and eject it vertically at that point.

In order to test the photogrammetry based trajectory dependence on computed debris density error from Fig. 8, a matrix of trajectories were generated. The assumption was that the surveyed points from Table 1 are exact, since we know that the differential GPS accuracy is better than 1 mm. It is then assumed that the source of all error is due to the placement of reference points and debris points in the image sequence set. This assumption implies that the algorithms described in the appendices and the image analysis software which uses those algorithms are not contributors to the final debris density error. Also implied in this assumption is that there are no mistakes made in identifying reference points in the images.  With this rather idealistic set of assumptions in mind, multiple trajectories were produced by deleting all reference points or all debris location points then reentering these points (a manual process) via the software graphical user interface.  The results from nine trajectories created in this way show that most of the variance is in the horizontal *xy*-plane, with much less error in the vertical direction.  As a result, the variation in debris density from the method of Fig. 8 is insignificant.  This provides a good level of confidence that the unknown debris density is unambiguous within the limits of the assumption of error sources.  Of course, if there are mistakes in one or more reference points, a significant change in debris density is possible.  It is also possible that uncertainties in the drag coefficient theory used in this analysis from Loth [5] could bias results.

## 6. Summary and Conclusions

The photogrammetry method described in this paper is based on a methodology that assumes that camera positions and orientations are unknown.   The software implementation of that method was previously developed for a space shuttle accident investigation.  Other variations of object tracking, in particular bird tracking over the launch pad, have recently been implemented at KSC.  In that case, coordinates of the cameras are determined from a differential GPS survey (this amounts to measuring a point on the ground, then parking a trailer containing the remote tracking mechanism and cameras over that point, then estimating the distance to the camera CCD or focal point). Once the camera positions are estimated, the number of reference points needed to solve the photogrammetry problem is dramatically reduced, and in the case of the bird tracking system, only two reference points are required [9].  In future work, it would be desirable to compare trajectories derived from both of these methods.

The method combining multicamera photogrammetry with ballistic analysis using the $R^2$ norm has successfully identified the high-flyer debris as low-density foam, most likely from the SRB throat plug, and not fire brick from beneath the Mobile Launch Platform. The method is valuable because it may be implemented as an automated process to locate and identify the sources of debris within seconds of liftoff. In this method, the photogrammetry plays the role of interpolating between nonsynchronized images from multiple cameras, triangulating the multiple two-dimensional images into a set of data points for the three-dimensional trajectory of the projectile, and measuring the projectile's physical dimensions. The ballistic analysis plays the role of regressing to a best estimate of the initial conditions and ballistic coefficient of the trajectory and thus providing a best estimate of the projectile's material density. Back-projecting the trajectory also indicates the projectile's point of origin, while the mass density and size provide constraints upon its possible composition and identity. These were found in the present example to be effective and sufficient. The combination of multicamera photogrammetry and automated ballistics analysis thus makes it possible to rapidly identify debris densities in future launch events. With further development of this technique, it may be possible to correlate projectile orientation as it spins with the instantaneous accelerations observed in the trajectory, including the crossrange motion, thus reducing the uncertainty in material composition and its physical dimensions. More advanced algorithms, possibly including artificial intelligence, may further improve the degree of automation, the accuracy, and the speed in identifying projectiles. This could make it possible to diagnose in near-real time the underlying root cause of debris-releasing events and how they may have affected the launch vehicle. This is analogous to the automated analysis that has been implemented at high-energy particle colliders, wherein a shower of particle trajectories is analyzed one trajectory at a time to identify the composition of each individual particle and thus piece together the physical conditions that emitted the overall shower. Analytical speed of this sort is important for launch pad environments when deciding whether possible debris damage has made it unsafe to perform critical flight activities such as landing the spacecraft.

**Acknowledgements**

The authors would like to thank Tom Ford of the NASA Image Analysis Lab at John F. Kennedy Space Center, Florida, and his team for providing the scanned digital images from the Launch Pad 39A perimeter high-speed film cameras. We are also indebted to Kevin Beuer of EG&E Tech and his team for providing a high-resolution survey of photogrammetry reference points.   The authors also wish to acknowledge the efforts of the ASRC Aerospace Engineering Support Group at KSC for their efforts in preparing this manuscript for submission.

# Appendix A

$$\nabla \mathbf{E} = 2\sum_{k=1}^{N} \{\mu_k(R_x x_k + R_y y_k + R_z z_k + 1) - (a_x x_k + b_x y_k + c_x z_k) - m_0\} \cdot \boldsymbol{\alpha}_k$$

$$+ 2\sum_{k=1}^{N} \{v_k(R_x x_k + R_y y_k + R_z z_k + 1) - (a_y x_k + b_y y_k + c_y z_k) - n_0\} \cdot \boldsymbol{\beta}_k \quad . \quad \text{(A-1)}$$

$$= 0$$

The vectors $\boldsymbol{\alpha}_k$ and $\boldsymbol{\beta}_k$ are the partial derivatives with respect to the parameter vector $\mathbf{P}$:

$$\mathbf{P} \equiv \begin{pmatrix} a_x \\ b_x \\ c_x \\ a_y \\ b_y \\ c_y \\ R_x \\ R_y \\ R_z \\ m_0 \\ n_0 \end{pmatrix} \quad \boldsymbol{\alpha}_k \equiv \begin{pmatrix} -x_k \\ -y_k \\ -z_k \\ 0 \\ 0 \\ 0 \\ \mu_k x_k \\ \mu_k y_k \\ \mu_k z_k \\ -1 \\ 0 \end{pmatrix} \quad \boldsymbol{\beta}_k \equiv \begin{pmatrix} 0 \\ 0 \\ 0 \\ -x_k \\ -y_k \\ -z_k \\ v_k x_k \\ v_k y_k \\ v_k z_k \\ 0 \\ -1 \end{pmatrix} \quad . \quad \text{(A-2)}$$

The gradient of $E$, Equation (A-1), can be more simply expressed as:

$$\nabla \mathbf{E} = 2(\boldsymbol{\Gamma} \cdot \mathbf{P} - \mathbf{A}) \quad , \quad \text{(A-3)}$$

where it can be shown that:

$$\boldsymbol{\Gamma} = \sum_{k=1}^{N} \{|\boldsymbol{\alpha}_k\rangle\langle\boldsymbol{\alpha}_k| + |\boldsymbol{\beta}_k\rangle\langle\boldsymbol{\beta}_k|\} \quad \text{and} \quad \text{(A-4a)}$$

$$\mathbf{A} = -\sum_{k=1}^{N}(\mu_k \boldsymbol{\alpha}_k + v_k \boldsymbol{\beta}_k) \quad . \quad \text{(A-4b)}$$

The matrix $\boldsymbol{\Gamma}$ and vector $\mathbf{A}$ are solely based on the reference point data. Therefore, setting the gradient of $E$ in Equation (A-3) to zero, leads to the following solution for the camera model parameters:

$$\mathbf{P} = \boldsymbol{\Gamma}^{-1} \cdot \mathbf{A} \quad . \quad \text{(A-5)}$$

Complete Expansion of Camera Model Equations (A-3) and (A-4a) and (A-4b):

$$\nabla \mathbf{E} = 2(\mathbf{\Gamma} \cdot \mathbf{P} - \mathbf{A})$$

$$= 2\sum_{k=1}^{N} \begin{pmatrix} -\mu_k(R_x x_k + R_y y_k + R_z z_k + 1)x_k + a_x x_k^2 + b_x x_k y_k + c_x x_k z_k + m_0 x_k \\ -\mu_k(R_x x_k + R_y y_k + R_z z_k + 1)y_k + a_x x_k y_k + b_x y_k^2 + c_x y_k z_k + m_0 y_k \\ -\mu_k(R_x x_k + R_y y_k + R_z z_k + 1)z_k + a_x x_k z_k + b_x y_k z_k + c_x z_k^2 + m_0 z_k \\ -\nu_k(R_x x_k + R_y y_k + R_z z_k + 1)x_k + a_y x_k^2 + b_y x_k y_k + c_y x_k z_k + n_0 x_k \\ -\nu_k(R_x x_k + R_y y_k + R_z z_k + 1)y_k + a_y x_k y_k + b_y y_k^2 + c_y y_k z_k + n_0 y_k \\ -\nu_k(R_x x_k + R_y y_k + R_z z_k + 1)z_k + a_y x_k z_k + b_y y_k z_k + c_y z_k^2 + n_0 z_k \\ (\mu_k^2 + \nu_k^2)(R_x x_k + R_y y_k + R_z z_k + 1)x_k - (a_x x_k^2 + b_x x_k y_k + c_x x_k z_k)\mu_k - (a_y x_k^2 + b_y x_k y_k + c_y x_k z_k)\nu_k - (m_0 \mu_k + n_0 \nu_k)x_k \\ (\mu_k^2 + \nu_k^2)(R_x x_k + R_y y_k + R_z z_k + 1)y_k - (a_x x_k y_k + b_x y_k^2 + c_x y_k z_k)\mu_k - (a_y x_k y_k + b_y y_k^2 + c_y y_k z_k)\nu_k - (m_0 \mu_k + n_0 \nu_k)y_k \\ (\mu_k^2 + \nu_k^2)(R_x x_k + R_y y_k + R_z z_k + 1)z_k - (a_x x_k z_k + b_x y_k z_k + c_x z_k^2)\mu_k - (a_y x_k z_k + b_y y_k z_k + c_y z_k^2)\nu_k - (m_0 \mu_k + n_0 \nu_k)z_k \\ -\mu_k(R_x x_k + R_y y_k + R_z z_k + 1) + a_x x_k + b_x y_k + c_x z_k + m_0 \\ -\nu_k(R_x x_k + R_y y_k + R_z z_k + 1) + a_y x_k + b_y y_k + c_y z_k + n_0 \end{pmatrix} \quad \text{(A-6)}$$

$$\mathbf{\Gamma} = \sum_{k=1}^{N}\{|\mathbf{\alpha}_k\rangle\langle\mathbf{\alpha}_k| + |\mathbf{\beta}_k\rangle\langle\mathbf{\beta}_k|\}$$

$$= \sum_{k=1}^{N} \begin{pmatrix} x_k^2 & x_k y_k & x_k z_k & 0 & 0 & 0 & -\mu_k x_k^2 & -\mu_k x_k y_k & -\mu_k x_k z_k & x_k & 0 \\ x_k y_k & y_k^2 & y_k z_k & 0 & 0 & 0 & -\mu_k x_k y_k & -\mu_k y_k^2 & -\mu_k y_k z_k & y_k & 0 \\ x_k z_k & y_k z_k & z_k^2 & 0 & 0 & 0 & -\mu_k x_k z_k & -\mu_k y_k z_k & -\mu_k z_k^2 & z_k & 0 \\ 0 & 0 & 0 & x_k^2 & x_k y_k & x_k z_k & -\nu_k x_k^2 & -\nu_k x_k y_k & -\nu_k x_k z_k & 0 & x_k \\ 0 & 0 & 0 & x_k y_k & y_k^2 & y_k z_k & -\nu_k x_k y_k & -\nu_k y_k^2 & -\nu_k y_k z_k & 0 & y_k \\ 0 & 0 & 0 & x_k z_k & y_k z_k & z_k^2 & -\nu_k x_k z_k & -\nu_k y_k z_k & -\nu_k z_k^2 & 0 & z_k \\ -\mu_k x_k^2 & -\mu_k x_k y_k & -\mu_k x_k z_k & -\nu_k x_k^2 & -\nu_k x_k y_k & -\nu_k x_k z_k & (\mu_k^2+\nu_k^2)x_k^2 & (\mu_k^2+\nu_k^2)x_k y_k & (\mu_k^2+\nu_k^2)x_k z_k & -\mu_k x_k & -\nu_k x_k \\ -\mu_k x_k y_k & -\mu_k y_k^2 & -\mu_k y_k z_k & -\nu_k x_k y_k & -\nu_k y_k^2 & -\nu_k y_k z_k & (\mu_k^2+\nu_k^2)x_k y_k & (\mu_k^2+\nu_k^2)y_k^2 & (\mu_k^2+\nu_k^2)y_k z_k & -\mu_k y_k & -\nu_k y_k \\ -\mu_k x_k z_k & -\mu_k y_k z_k & -\mu_k z_k^2 & -\nu_k x_k z_k & -\nu_k y_k z_k & -\nu_k z_k^2 & (\mu_k^2+\nu_k^2)x_k z_k & (\mu_k^2+\nu_k^2)y_k z_k & (\mu_k^2+\nu_k^2)z_k^2 & -\mu_k z_k & -\nu_k z_k \\ x_k & y_k & z_k & 0 & 0 & 0 & -\mu_k x_k & -\mu_k y_k & -\mu_k z_k & 1 & 0 \\ 0 & 0 & 0 & x_k & y_k & z_k & -\nu_k x_k & -\nu_k y_k & -\nu_k z_k & 0 & 1 \end{pmatrix} \quad \text{(A-7)}$$

$$\mathbf{A} = -\sum_{k=1}^{N} \mu_k \mathbf{\alpha}_k + \nu_k \mathbf{\beta}_k$$

$$= \sum_{k=1}^{N} \begin{pmatrix} \mu_k x_k \\ \mu_k y_k \\ \mu_k z_k \\ \nu_k x_k \\ \nu_k y_k \\ \nu_k z_k \\ -(\mu_k^2 + \nu_k^2)x_k \\ -(\mu_k^2 + \nu_k^2)y_k \\ -(\mu_k^2 + \nu_k^2)z_k \\ \mu_k \\ \nu_k \end{pmatrix}. \quad \text{(A-8)}$$

# Appendix B

$$\nabla \mathbf{E} = 2\sum_{j=1}^{M}\left\{\mu_j\left(R_{x_j}x + R_{y_j}y + R_{z_j}z + 1\right) - \left(a_{x_j}x + b_{x_j}y + c_{x_j}z\right) - m_{0_j}\right\}\cdot \boldsymbol{\alpha}_j$$

$$+ 2\sum_{k=1}^{N}\left\{v_{k_j}\left(R_{x_j}x + R_{y_j}y + R_{z_j}z + 1\right) - \left(a_{y_j}x + b_{y_j}y + c_{y_j}z\right) - n_{0_j}\right\}\cdot \boldsymbol{\beta}_j \quad \text{(B-1)}$$

$$= 0$$

The vectors $\boldsymbol{\alpha}_j$ and $\boldsymbol{\beta}_j$ are the partial derivatives with respect to the parameter vector $\mathbf{P}$:

$$\mathbf{P} \equiv \begin{pmatrix} x \\ y \\ z \end{pmatrix} \qquad \boldsymbol{\alpha}_j \equiv \begin{pmatrix} \mu_j R_{x_j} - a_{x_j} \\ \mu_j R_{y_j} - b_{x_j} \\ \mu_j R_{z_j} - c_{x_j} \end{pmatrix} \qquad \boldsymbol{\beta}_j \equiv \begin{pmatrix} v_j R_{x_j} - a_{y_j} \\ v_j R_{y_j} - b_{y_j} \\ v_j R_{z_j} - c_{y_j} \end{pmatrix}. \quad \text{(B-2)}$$

The gradient of $E$, Equation (B-1), can be expressed as:

$$\nabla \mathbf{E} = 2(\boldsymbol{\Gamma}\cdot \mathbf{P} - \mathbf{A}), \quad \text{(B-3)}$$

where it can be shown that:

$$\boldsymbol{\Gamma} = \sum_{j=1}^{M}\left\{|\boldsymbol{\alpha}_j\rangle\langle\boldsymbol{\alpha}_j| + |\boldsymbol{\beta}_j\rangle\langle\boldsymbol{\beta}_j|\right\} \quad \text{and} \quad \mathbf{A} = -\sum_{j=1}^{M}\left(\mu_j \boldsymbol{\alpha}_j + v_j \boldsymbol{\beta}_j\right), \quad \text{(B-4)}$$

and again, the outer product of $\boldsymbol{\alpha}_j$ with itself, plus the outer product of $\boldsymbol{\beta}_j$ with itself, summed over all $M$ cameras, yields $\boldsymbol{\Gamma}$. Setting the gradient of $E$ in Equation (B-3) to zero, leads to the following solution for the unknown position coordinate:

$$\mathbf{P} = \boldsymbol{\Gamma}^{-1}\cdot \mathbf{A} = \begin{pmatrix} x \\ y \\ z \end{pmatrix}. \quad \text{(B-5)}$$

Complete Expansion of Camera Model Equations (B-3) and (B-4):

$$\nabla \mathbf{E} = 2(\mathbf{\Gamma} \cdot \mathbf{P} - \mathbf{A})$$

$$= 2\sum_{j=1}^{M} \begin{pmatrix} \{\mu_j(R_{x_j}x + R_{y_j}y + R_{z_j}z + 1) - (a_{x_j}x + b_{x_j}y + c_x z) - m_{0_j}\}(\mu_j R_{x_j} - a_{x_j}) \\ \{\mu_j(R_{x_j}x + R_{y_j}y + R_{z_j}z + 1) - (a_{x_j}x + b_{x_j}y + c_x z) - m_{0_j}\}(\mu_j R_{y_j} - b_{x_j}) \\ \{\mu_j(R_{x_j}x + R_{y_j}y + R_{z_j}z + 1) - (a_{x_j}x + b_{x_j}y + c_x z) - m_{0_j}\}(\mu_j R_{z_j} - c_{x_j}) \end{pmatrix}$$

$$+ 2\sum_{j=1}^{M} \begin{pmatrix} \{\nu_{k_j}(R_{x_j}x + R_{y_j}y + R_{z_j}z + 1) - (a_{y_j}x + b_{y_j}y + c_{y_j}z) - n_{0_j}\}(\nu_j R_{x_j} - a_{y_j}) \\ \{\nu_{k_j}(R_{x_j}x + R_{y_j}y + R_{z_j}z + 1) - (a_{y_j}x + b_{y_j}y + c_{y_j}z) - n_{0_j}\}(\nu_j R_{y_j} - b_{y_j}) \\ \{\nu_{k_j}(R_{x_j}x + R_{y_j}y + R_{z_j}z + 1) - (a_{y_j}x + b_{y_j}y + c_{y_j}z) - n_{0_j}\}(\nu_j R_{z_j} - c_{y_j}) \end{pmatrix} \quad . \quad \text{(B-6)}$$

$$\mathbf{\Gamma} = \sum_{j=1}^{M} \{|\boldsymbol{\alpha}_j\rangle\langle\boldsymbol{\alpha}_j| + |\boldsymbol{\beta}_j\rangle\langle\boldsymbol{\beta}_j|\}$$

$$= \sum_{j=1}^{M} \begin{pmatrix} (\mu_j R_{x_j} - a_{x_j})(\mu_j R_{x_j} - a_{x_j}) & (\mu_j R_{x_j} - a_{x_j})(\mu_j R_{y_j} - b_{x_j}) & (\mu_j R_{x_j} - a_{x_j})(\mu_j R_{z_j} - c_{x_j}) \\ (\mu_j R_{y_j} - b_{x_j})(\mu_j R_{x_j} - a_{x_j}) & (\mu_j R_{y_j} - b_{x_j})(\mu_j R_{y_j} - b_{x_j}) & (\mu_j R_{y_j} - b_{x_j})(\mu_j R_{z_j} - c_{x_j}) \\ (\mu_j R_{z_j} - c_{x_j})(\mu_j R_{x_j} - a_{x_j}) & (\mu_j R_{z_j} - c_{x_j})(\mu_j R_{y_j} - b_{x_j}) & (\mu_j R_{z_j} - c_{x_j})(\mu_j R_{z_j} - c_{x_j}) \end{pmatrix} \quad . \quad \text{(B-7)}$$

$$+ \sum_{j=1}^{M} \begin{pmatrix} (\nu_j R_{x_j} - a_{y_j})(\nu_j R_{x_j} - a_{y_j}) & (\nu_j R_{x_j} - a_{y_j})(\nu_j R_{y_j} - b_{y_j}) & (\nu_j R_{x_j} - a_{y_j})(\nu_j R_{z_j} - c_{y_j}) \\ (\nu_j R_{y_j} - b_{y_j})(\nu_j R_{x_j} - a_{y_j}) & (\nu_j R_{y_j} - b_{y_j})(\nu_j R_{y_j} - b_{y_j}) & (\nu_j R_{y_j} - b_{y_j})(\nu_j R_{z_j} - c_{y_j}) \\ (\nu_j R_{z_j} - c_{y_j})(\nu_j R_{x_j} - a_{y_j}) & (\nu_j R_{z_j} - c_{y_j})(\nu_j R_{y_j} - b_{y_j}) & (\nu_j R_{z_j} - c_{y_j})(\nu_j R_{z_j} - c_{y_j}) \end{pmatrix}$$

$$\mathbf{A} = -\sum_{j=1}^{M} \mu_j \boldsymbol{\alpha}_j + \nu_j \boldsymbol{\beta}_j$$

$$= \sum_{j=1}^{M} \begin{pmatrix} (\mu_j R_{x_j} - a_{x_j})(m_{0_j} - \mu_j) + (\nu_j R_{x_j} - a_{y_j})(n_{0_j} - \nu_j) \\ (\mu_j R_{y_j} - b_{x_j})(m_{0_j} - \mu_j) + (\nu_j R_{y_j} - b_{y_j})(n_{0_j} - \nu_j) \\ (\mu_j R_{z_j} - c_{x_j})(m_{0_j} - \mu_j) + (\nu_j R_{z_j} - c_{y_j})(n_{0_j} - \nu_j) \end{pmatrix} \quad . \quad \text{(B-8)}$$

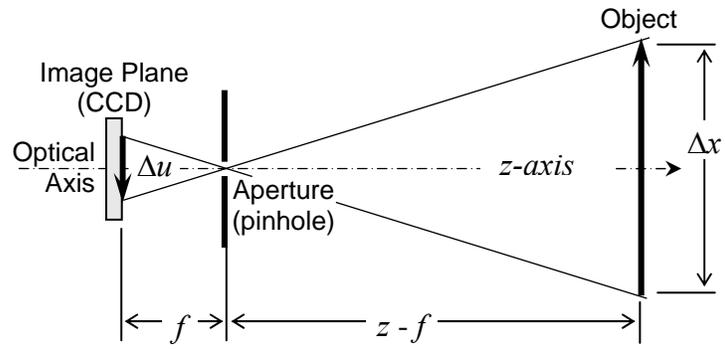

Fig. 1. Pinhole camera model (PCM) showing definition of focal length, $f$.

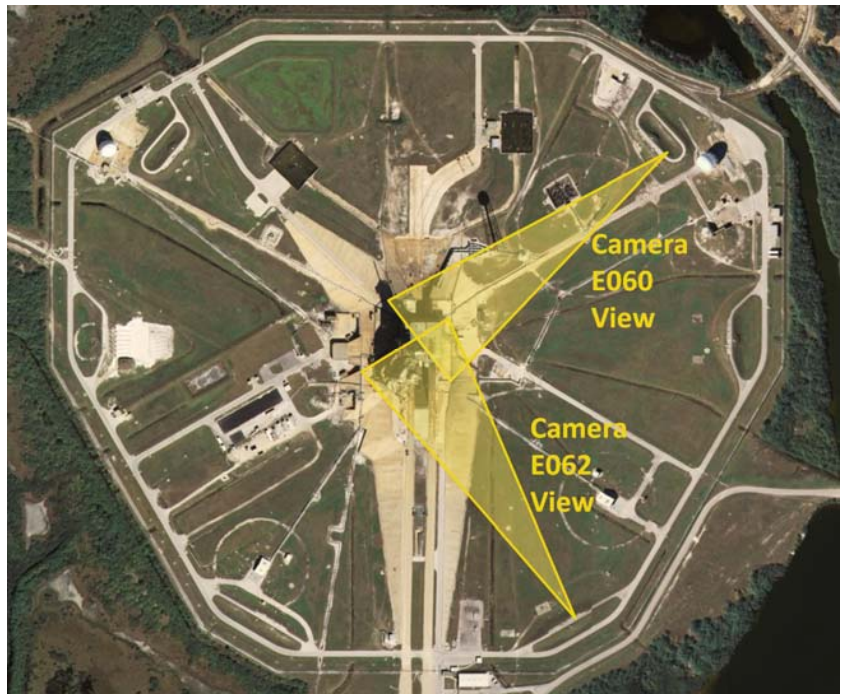

Fig. 2. Camera E060 and Camera E062 view.

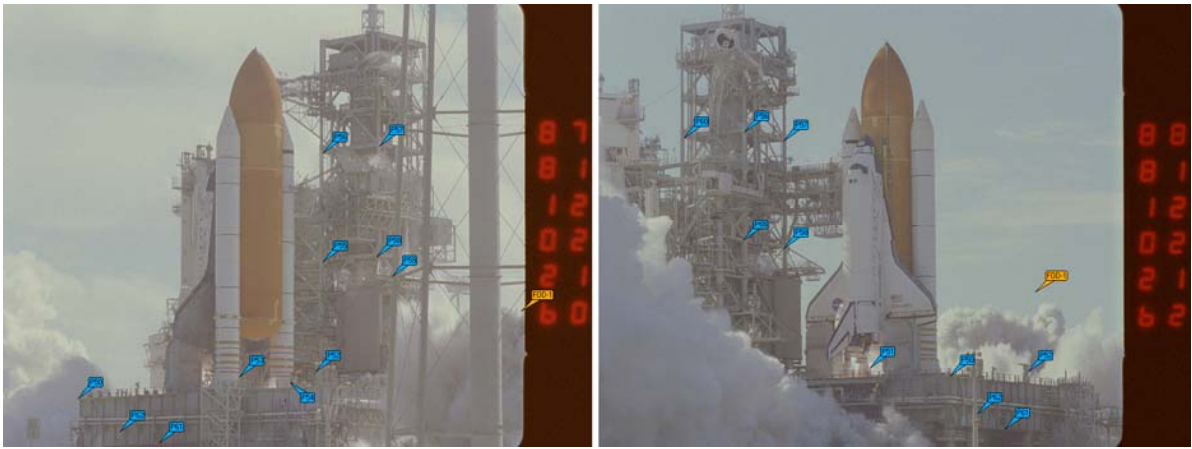

Fig. 3. Pad 39A structure, showing reference points that were actually used: left is view from camera E060; right is from camera E062.

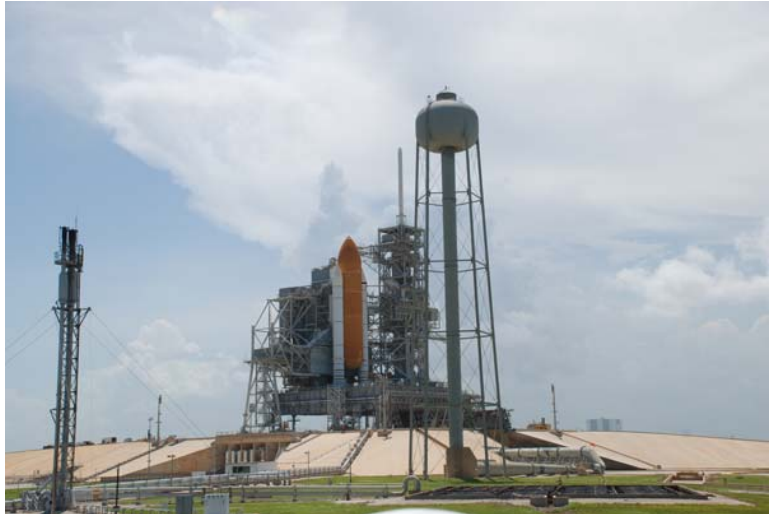
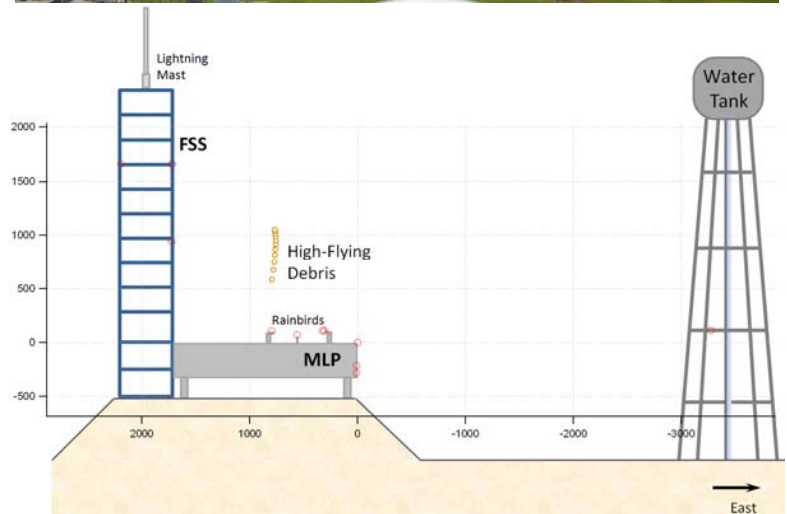
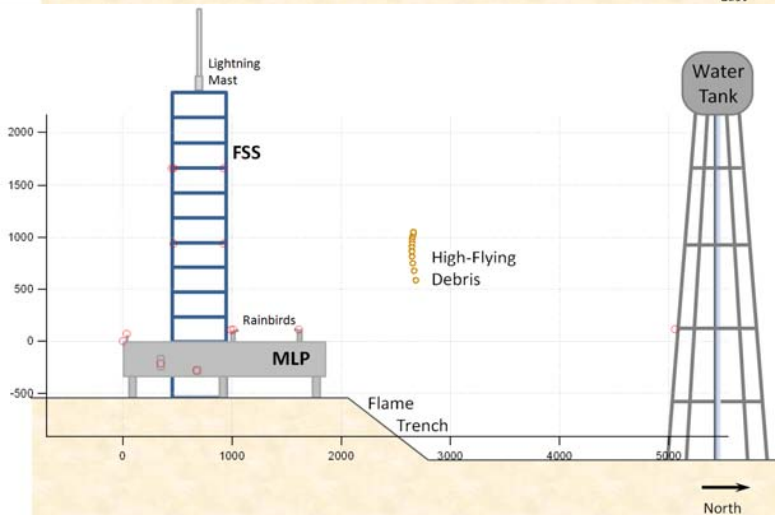

Fig. 4. Top is view from camera E060 site; middle is a plot showing reference points and debris track from a north-looking view; bottom is a plot showing reference points and debris track from a west-looking view. Axes show coordinates (in inches) with southesast corner of MLP as origin.

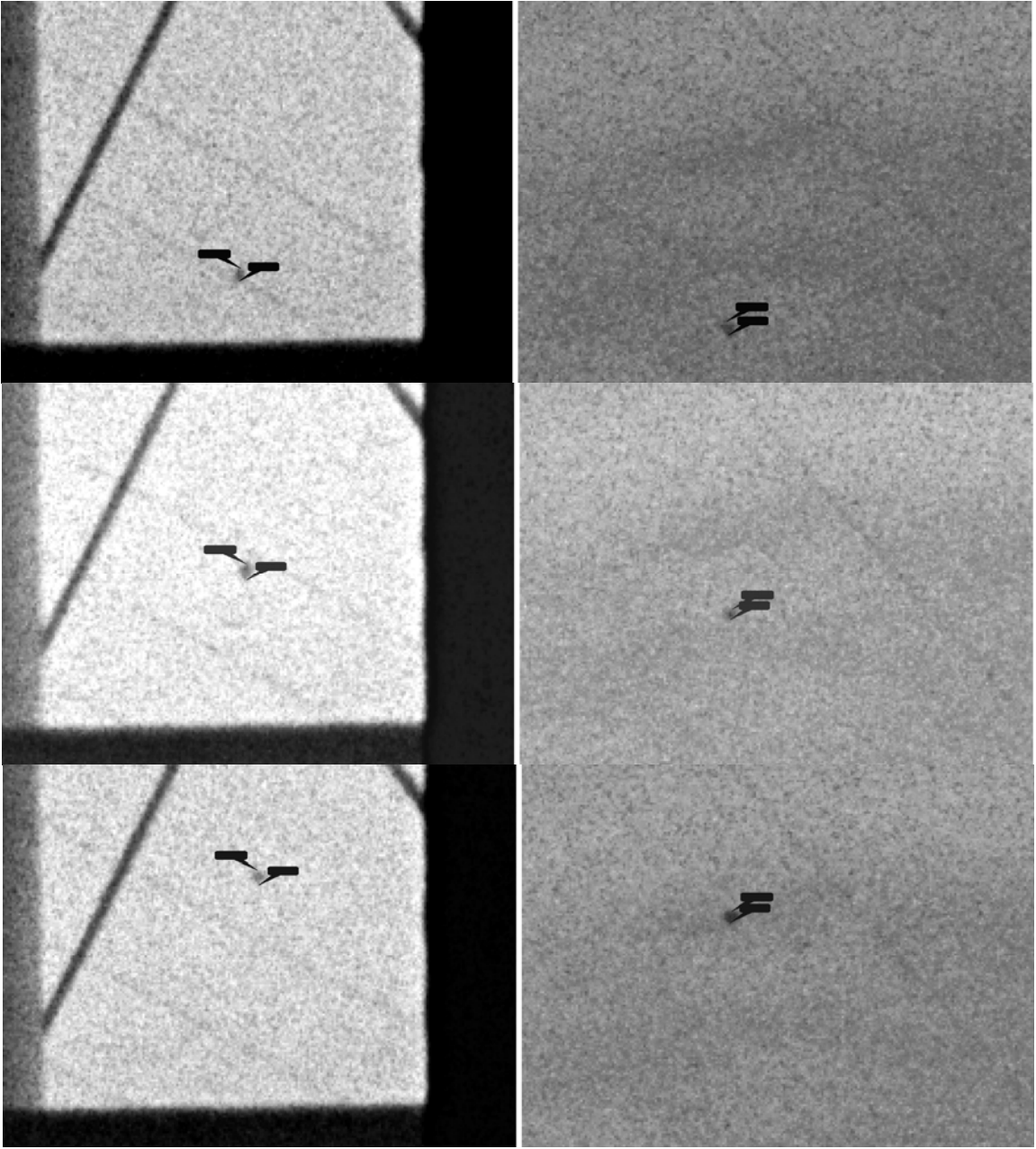

Fig. 5. Close-up of debris observed from cameras E060 (left) and E062 (right). Top is relative time 13.165 s; middle is time 13.220 s; and bottom is time 13.276 s. Note that this sequence corresponds to 18 fps, by using a decimation factor of 10 (both cameras run at 180 fps).

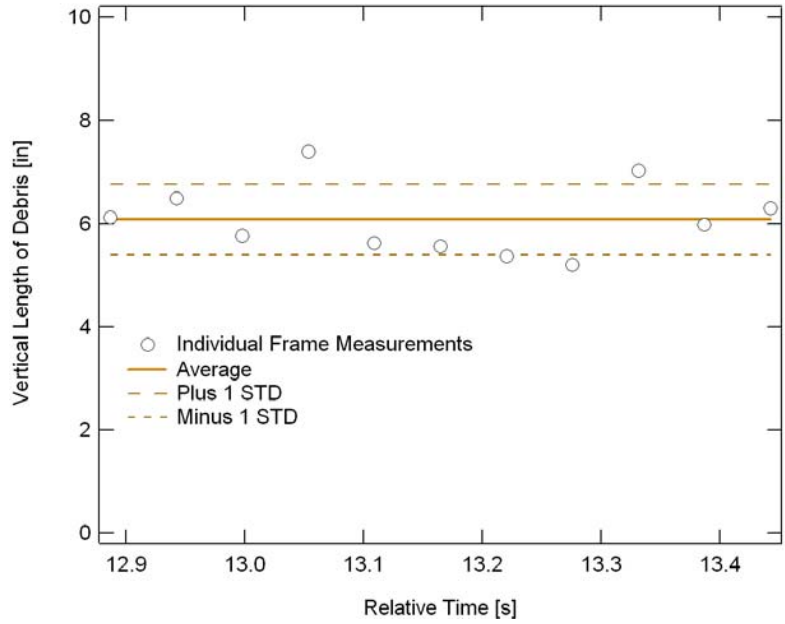

Fig. 6. Debris size estimate and standard deviation of mean.

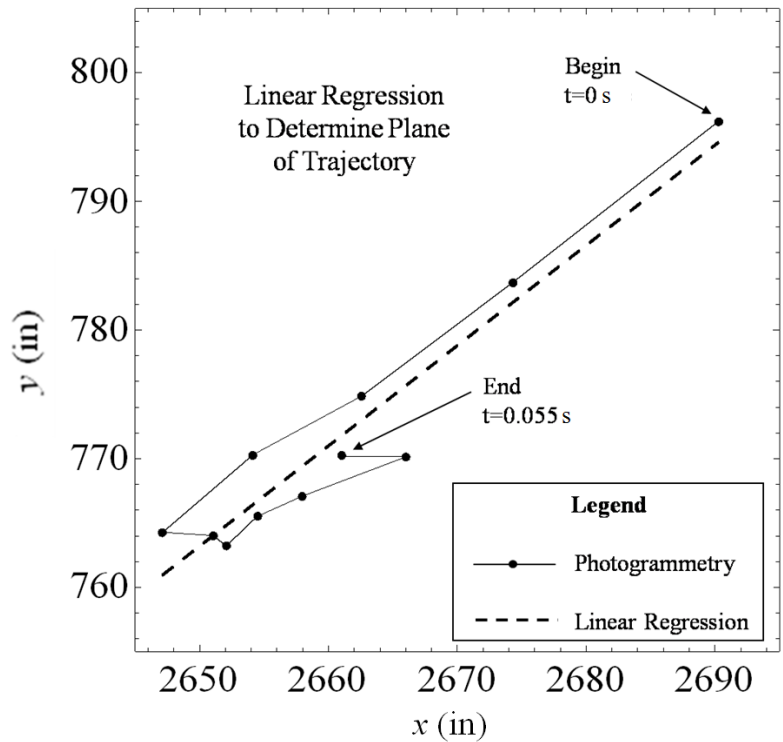

Fig. 7. Linear regression of trajectory points looking downward into the *xy* plane.

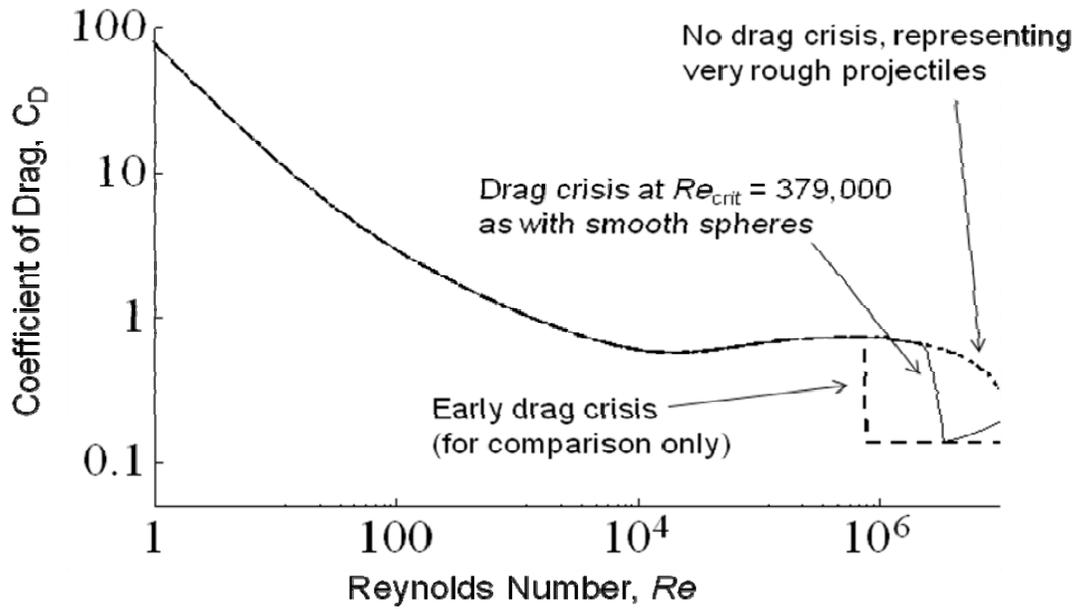

Fig. 8. Coefficient of Drag as a function of Reynolds number, using the Knudsen number *Kn* for the ambient atmosphere sea level on the day of launch. Mach number is a function of *Re* and *Kn*. The solid line follows Loth [5] but multiplied by a shape factor of 1.5.

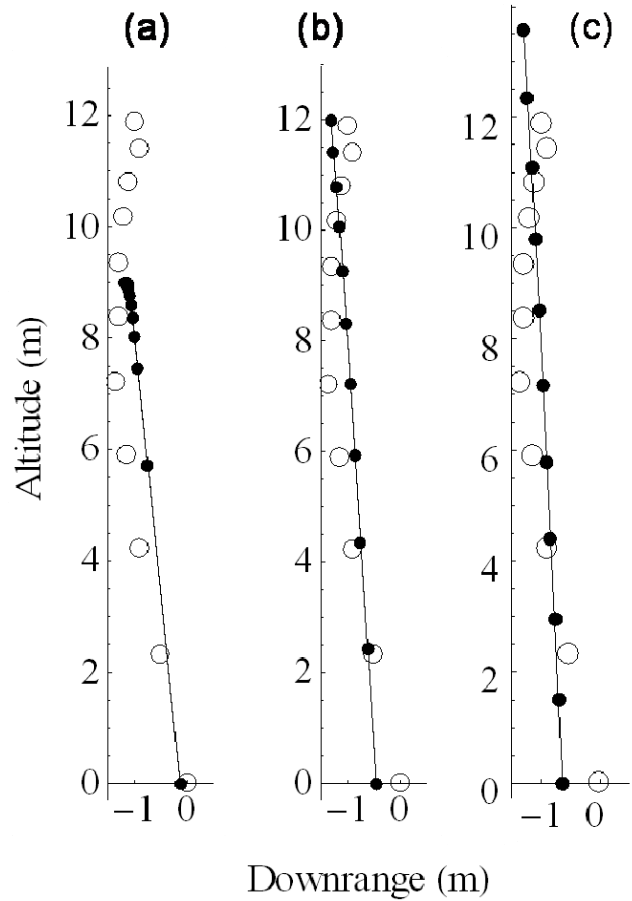

Fig. 9. Solid lines and dots: calculated trajectories with least square fits for three specified projectile densities: (a) 5.99 kg/m$^3$, (b) 52.22 kg/m$^3$ (the best fit), and (c) 26,700 kg/m$^3$, illustrating how deceleration depends sensitively on projectile density. Open circles: photogrammetry data at identical times as the solid dots. The origin is set at the location of the first photogrammetry point, not at ground level of the launch pad. Note the sizes of the markers are not indicators of error magnitude.

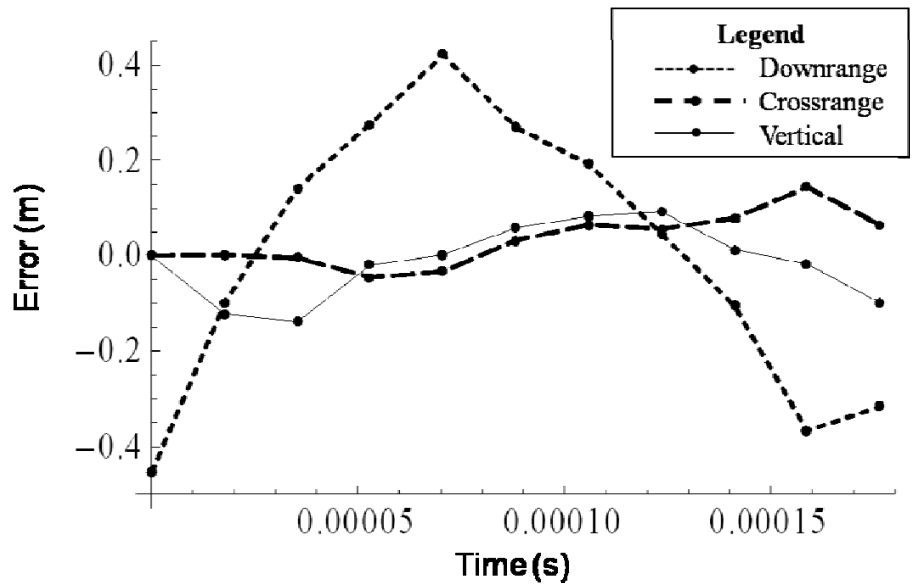

Fig. 10. Differences between numerical trajectory and photogrammetry data.

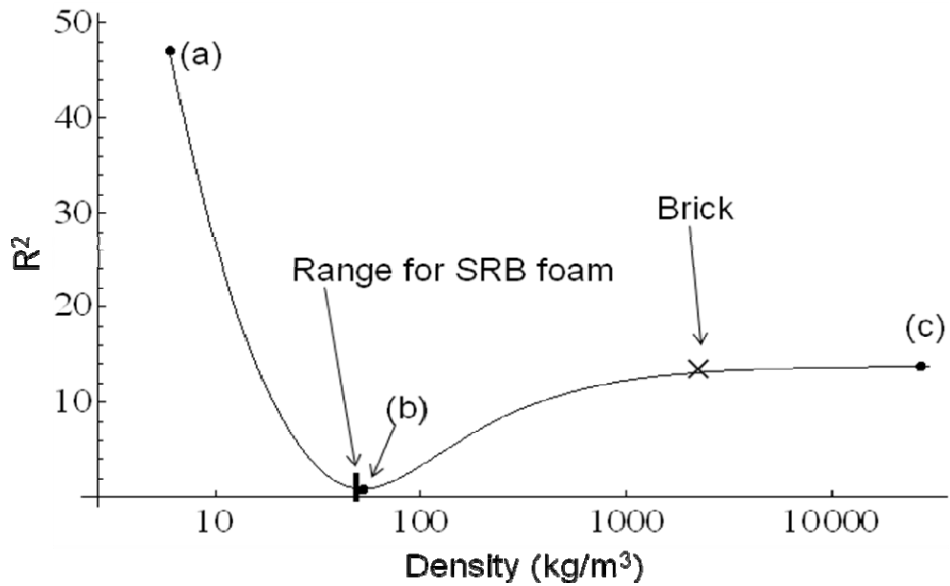

Fig. 11. $R^2$ norm versus density with shape factor $C_S = 1.5$. The three cases shown in Figure 9 are indicated by solid dots: (a) 5.99 kg/m$^3$, (b) 52.22 kg/m$^3$ (the best fit), and (c) 26,700 kg/m$^3$. The thickness of the vertical bar indicates the range of density values for SRB foam. The × indicates the density value for flame trench brick.

Table 1. Surveyed photogrammetry reference points.

| Ref Pt | $x$ (in) | $y$ (in) | $z$ (in) |
|---|---|---|---|
| P50 | 0 | 0 | 0 |
| P51 | 36.06 | 561.324 | 70.968 |
| P52 | 1611.06 | 309.48 | 110.592 |
| P53 | 1012.08 | 323.964 | 109.356 |
| P54 | 990.864 | 795.336 | 107.628 |
| P55 | 465.564 | 1722.084 | 941.4 |
| P56 | 465.048 | 1720.908 | 1661.172 |
| P57 | 923.448 | 1722.792 | 1660.896 |
| P58 | 922.536 | 1723.428 | 941.616 |
| P59 | 5056.452 | -3272.52 | 113.844 |
| P60 | 448.188 | 2194.644 | 1661.04 |
| P61 | 675.756 | 9 | -281.94 |
| P62 | 346.428 | 8.436 | -214.32 |

Table 2. Trajectory data points obtained from photogrammetry.

| Time (s) | x (in)   | y (in)   | z (in)   |
|----------|----------|----------|----------|
| 0        | 2690.226 | 796.1924 | 585.2793 |
| 0.005    | 2674.273 | 783.6768 | 676.6074 |
| 0.011    | 2662.621 | 774.8918 | 751.6152 |
| 0.016    | 2654.143 | 770.3081 | 817.6101 |
| 0.022    | 2647.139 | 764.281  | 869.4896 |
| 0.027    | 2651.044 | 764.022  | 914.8872 |
| 0.033    | 2652.058 | 763.208  | 952.9451 |
| 0.039    | 2654.459 | 765.5084 | 985.555  |
| 0.044    | 2657.936 | 767.0779 | 1010.74  |
| 0.05     | 2666.022 | 770.1479 | 1034.449 |
| 0.055    | 2661.027 | 770.1991 | 1053.326 |

Table 3.  In-plane trajectory data.

| Time (s) | Downrange (m) | Altitude (m) |
|---|---|---|
| 0 | 0 | 0 |
| 0.005 | 0.535518 | 1.88924 |
| 0.011 | 1.09381 | 3.55889 |
| 0.016 | 1.52688 | 5.08732 |
| 0.022 | 2.19606 | 6.2563 |
| 0.027 | 2.71402 | 7.31281 |
| 0.033 | 3.16113 | 8.22921 |
| 0.039 | 3.61244 | 8.94424 |
| 0.044 | 4.13373 | 9.62042 |
| 0.05 | 4.44928 | 10.1258 |